# Retrieval-Augmented Large Language Models for Evidence-Informed Guidance on Cannabidiol Use in Older Adults

Ali Abedi, Ph.D.,[1,2*], Charlene H. Chu, Ph.D.,[1,2], Shehroz S. Khan, Ph.D.,[3]

[1] Lawrence Bloomberg Faculty of Nursing, University of Toronto, Toronto, Canada
[2] KITE Research Institute, Toronto Rehabilitation Institute, University Health Network, Toronto, ON, Canada
[3] College of Engineering and Technology, American University of the Middle East, Kuwait
*ali.abedi@uhn.ca

## Abstract

**Background:** Older adults often experience chronic conditions such as pain and sleep disturbances, leading many to explore cannabidiol (CBD) for symptom relief. Safe use requires appropriate dosing, careful titration, and awareness of potential interactions; however, the stigma surrounding CBD use and limited health literacy can constrain comprehension. Conversational AI systems built on large language models (LLMs) and retrieval-augmented generation (RAG) may support CBD education, but their safety and reliability remain under-evaluated.

**Objective:** This study aimed to (1) design a retrieval-augmented LLM framework integrating structured prompts and evidence retrieval from curated CBD resources to generate safe, coherent, context-aware guidance for older adults, and (2) systematically evaluate leading LLMs and RAG systems using an automated, annotation-free framework in the absence of standardized benchmarks.

**Methods:** A structured parametric scenario generation framework produced sixty-four diverse profiles by varying symptom goals, administration preferences, cognitive status, demographics, health parameters, comorbidities, medication regimens, cannabis history, and caregiver support. These scenarios, combined with advanced prompt engineering, were used to test OpenAI GPT 5.1, Google Gemini 2.5 Pro, Mistral AI Medium 3, Anthropic Claude Sonnet 4.5, xAI Grok 4, and DeepSeek V3.2-Exp. Retrieval-augmented variants of GPT 5.1 and Gemini 2.5 Pro incorporated thirty-two curated CBD guidelines. A novel ensemble retrieval architecture was also designed, combining two independent RAG systems with a third tiebreaker RAG. Model outputs were evaluated using three automated, annotation-free methods proposed in this study, including statistical consensus analysis, feature-aligned directional checks, and LLM-as-a-judge rubric scoring.

**Results:** Across all three evaluation strategies, the ensemble RAG configuration produced the most cautious, clinically grounded recommendations, closely followed by Gemini 2.5 Pro RAG and GPT 5.1 RAG. Standalone GPT 5.1 and Gemini 2.5 Pro performed reliably but were less consistently cautious than their retrieval-augmented counterparts. DeepSeek V3.2-Exp and Grok 4 generated higher, more variable dosing patterns, while Medium 3 showed intermediate behavior with moderate safety and alignment, closely matching guideline-based dosage and titration.

**Conclusions:** This study introduces a reproducible, annotation-free framework for benchmarking LLM-based CBD education and shows that retrieval-augmented models provide more adaptive guidance for older adults with diverse cognitive and clinical needs. The findings highlight the potential of structured retrieval to improve the reliability of AI-driven evidence-informed educational tools used in sensitive health contexts.

**Introduction**
Cannabidiol (CBD) is increasingly used by older adults to manage chronic pain [1], sleep disturbance [2], anxiety [3], and behavioral symptoms associated with neurocognitive disorders [4], [5]. Unlike tetrahydrocannabinol (THC), CBD is non-intoxicating [6] and has shown potential benefits for improving comfort, sleep quality, and overall well-being in later life [1], [2], [3], [4], [5]. Older adults are more likely to use cannabis for medical purposes than younger people [7], [8], [9]. Despite this growing interest, uncertainty regarding optimal dosing, titration, and drug interactions continues to raise safety concerns [10]. Older adults often have altered pharmacokinetics, polypharmacy, and heightened medication sensitivity, which increase the risk of adverse outcomes [10], [11]. These challenges are further magnified in individuals with mild cognitive impairment, who can experience difficulties processing written or verbal medication instructions [12]. Consequently, there is an urgent need for accessible, personalized, and evidence-grounded educational tools that can communicate CBD usage guidance safely and clearly to older adults and their caregivers.

Existing patient education materials, including clinical leaflets, online resources, and consumer pamphlets, often rely on static, text-heavy content that lacks personalization and contextual adaptation [13]. Literature also shows that older adults rely predominantly on online sources but can become overwhelmed with the amount of information, and subsequently, turn to friends for advice [14]. For individuals with cognitive changes or limited health literacy, such materials may be difficult to navigate or interpret [15], [16]. Advances in conversational artificial intelligence (AI) offer a potential pathway toward more adaptive and personalized health communication. Large language models (LLMs) are large-scale neural networks trained on extensive text corpora to recognize patterns in language [15], [16]. They apply this learned knowledge to generate responses and perform reasoning, enabling them to produce explanations, summaries, and recommendations tailored to user input [17]. Although LLMs can generate coherent and context-aware text, when used without domain-specific grounding, they may still produce incomplete, inaccurate, or misleading information [18]. In particular, LLMs are susceptible to hallucinations, whereby outputs may appear plausible but are not supported by factual evidence, potentially leading users to accept incorrect or non-evidence-based information [19]. These limitations are particularly concerning for medication-related education, where factual precision and alignment with safety guidance are essential [20], [21], [22], [23].

Retrieval-augmented generation (RAG) [24] enhances an LLM by retrieving relevant external documents and using them to augment the user's query before the model generates a response. By grounding the model in this augmented evidence, RAG can improve factual accuracy, consistency, and safety in the resulting output. LLMs such as OpenAI GPT 5.1 [25], Google Gemini 2.5 Pro [26], and Anthropic Claude Sonnet 4.5 [27] represent the newest generation of models capable of multimodal reasoning, with retrieval typically implemented through an external pipeline that supplies documents for synthesis. Despite this progress, previous studies in medication communication continue to report unreliable retrieval, limited transparency, inconsistent evaluation metrics, and a lack of standardized benchmarks for assessing educational dialogue quality [20], [21], [22], [23], [28]. In addition, to our knowledge, no existing work has evaluated RAG or LLM systems for CBD-related education or for older adults and those with cognitive impairment [20], [21], [22], [23], highlighting a clear methodological and clinical gap. This gap is particularly consequential given that cannabis use has increased more rapidly among older adults than any other age group and, despite growing normalization of medical cannabis, stigma continues to shape information-seeking [14]. In the absence of informed dialogue within formal care contexts, older adults may turn to recreational or non-medical sources [14], underscoring a critical methodological and clinical gap and the need to rigorously evaluate AI-enabled educational tools.

The primary goal of this study is to design and evaluate a retrieval-augmented LLM framework capable of delivering accurate, safe, and comprehensible educational communication on CBD use for older adults, including those with cognitive impairment. The framework addresses gaps in patient-centered medication education by generating evidence-based, context-aware explanations tailored to diverse clinical and cognitive profiles. Its objectives are twofold: (1) to develop a retrieval-augmented framework that integrates structured prompt engineering with evidence retrieval from curated CBD resources to produce safe, coherent, and context-aware communication for older adults, and (2) to systematically evaluate leading LLMs and RAG configurations across simulated scenarios using an automated, annotation-free assessment framework in the absence of standardized benchmarks or human reference labels.

This work introduces three core methodological contributions. First, it develops a retrieval-augmented LLM framework to generate evidence-grounded CBD education for older adults. Second, it implements a parametric scenario generation approach that models user variability in symptom goals, cognitive status, and health conditions to support reproducible evaluation. Third, it establishes an annotation-free evaluation pipeline that integrates consensus statistics, feature-aligned directional analysis, and LLM-as-a-judge rubric scoring to enable scalable and objective assessment of LLM and RAG systems in health communication.

The remainder of the paper is organized as follows. The next subsection reviews related work on the use of LLMs and RAG systems for medication communication. The subsequent section describes the methodological framework, including scenario design, prompt engineering, and evaluation procedures. This is followed by a presentation of the empirical results, and a final section that discusses the implications of the findings, key limitations, and directions for future research.

*Literature Review*

This subsection reviews the literature on LLM and RAG-based systems and conversational agents [29] in healthcare education, with attention to how these systems communicate medication information, dosing concepts, and safety considerations. It highlights both the potential of these technologies to improve access to clear health guidance and the limitations that may affect their suitability for older adults and individuals with cognitive changes.

A systematic review by Huo et al. [20] highlighted that treatment guidance was one of the most common applications of LLM-based conversational agents, including communication about medications and treatment options. The authors noted that published evaluations frequently lacked core methodological information, such as the specific model version, prompting strategy, or grounding requirements, making it difficult to assess or reproduce reported performance. They also emphasized that discussions of safety safeguards, error management, and ethical considerations were uncommon, and that almost no work addressed how these tools should be adapted for users with varying health literacy. The review by Emile et al. [21] similarly found that LLM-generated explanations generally aligned with clinical recommendations and were often rated as appropriate or clinically sound across diverse medical topics. However, they reported that the language used tended to exceed recommended patient reading levels, reducing accessibility for many users.

Amugongo et al. [22] systematically reviewed RAG for LLMs in healthcare and found that retrieval augmentation improves factual grounding, reduces hallucinations [18], and enhances clinical relevance compared with standalone LLMs. However, the authors also identified major gaps related to unreliable retrieval, limited transparency, privacy risks, and the absence of validated evaluation standards. In another review, Bunnell et al. [23] mapped RAG applications across health professions and highlighted ongoing concerns regarding bias, explainability, and overreliance on AI-generated recommendations. None of these reviews [20], [21], [22], [23] examined how such tools perform for older adults, including those with cognitive impairment, how caregivers may be integrated into their use, or how RAG and LLM systems address CBD-related education or dosing guidance, leaving a clear gap in the literature.

PharmaLLM [30] applied parameter-efficient fine-tuning to the Large Language Model Meta AI (Llama) 2 architecture using the Low-Rank Adaptation (LoRA) technique to create a domain-adapted medical prescription assistant with both text and speech capabilities. It was trained on medical data to improve accuracy and reduce unsafe outputs common in generic models. The system demonstrated strong overall performance across classification and usability metrics, indicating reliable response quality and user-friendly interaction. A user study reported that participants found the tool easy to navigate and its guidance relevant. Limitations included modest specificity in nuanced clinical cases and the use of an older backbone architecture, which could limit generalization and robustness compared with newer LLMs.

Cornelison et al. [31] evaluated the accuracy and completeness of GPT-3.5 in answering patient-centered medication questions derived from the Agency for Healthcare Research and Quality. Twelve standardized questions were posed for the top twenty prescribed drugs identified from the Medical Expenditure Panel Survey, generating 240 total responses. Two independent reviewers rated each answer on six-point and three-point scales for accuracy and completeness, respectively, with consensus scoring used to resolve discrepancies. The chatbot achieved high accuracy and completeness, with most responses rated correct and clinically acceptable. However, some outputs lacked essential patient guidance or contained partial inaccuracies. The authors emphasized that while GPT-3.5 can provide generally reliable medication information, human verification remains essential to ensure clinical safety and completeness.

Ismail et al. [32] developed a RAG system integrated with multilingual translation to improve drug information access and comprehension in low-resource healthcare settings. The model combined OpenAI's GPT-3.5-turbo with a Chroma vector database built from leading Nigerian medical information sources. Drug-related queries were processed through RAG, translated into Yoruba, Igbo, and Hausa, and evaluated by licensed pharmacists. The evaluation used twenty domain-specific prompts per language to assess accuracy, completeness, and linguistic quality. Results demonstrated strong translation performance, particularly in Yoruba and Hausa. Limitations included variability in translation quality across languages and the need for further fine-tuning on domain-specific datasets.

Med-Pal [33] was developed by fine-tuning several lightweight open-source LLMs on an expert-curated medication-enquiry dataset. All models were trained under a unified LoRA-based setup and evaluated using SCORE, a clinician-led rubric that rated safety, clinical accuracy, objectivity, reproducibility, and ease of understanding on a structured three-point scale. Mistral-7B [34] achieved the highest overall ratings and was selected as the Med-Pal backbone, then strengthened with safety guardrails. Med-Pal outperformed comparable lightweight biomedical models, although it still showed inconsistent reproducibility across repeated queries, limited coverage of highly complex or clinically nuanced medication scenarios, and dependence on an earlier-generation backbone, Mistral-7B, which could restrict generalization and robustness compared with newer architectures.

De Jesus et al. [35] used a dataset of 119 simulated outpatient prescription scenarios to evaluate Llama 3 across three configurations: a standard prompt, a structured prompt with explicit guidance, and a retrieval-augmented approach incorporating patient information leaflets from the Brazilian Health Regulatory Agency. Five physicians independently assessed the generated instructions for adequacy, clarity, personalization, and overall quality, while cosine similarity was computed against reference texts. The retrieval-augmented configuration produced the most accurate and complete instructions with fewer critical errors. The authors acknowledged that physician-based scoring introduced subjectivity despite predefined criteria and that simulated scenarios may not capture the complexity and diversity of real clinical practice, limiting generalizability. They further noted that the exclusion of special populations such as children, individuals with special needs, and those with polypharmacy may also constrain the model's applicability to broader patient groups.

Steybe et al. [36] developed GuideGPT, a context-aware chatbot using RAG integrated with a database of 449 scientific publications to answer clinical questions related to the prevention, diagnosis, and treatment of medication-related osteonecrosis of the jaw. Built on GPT-4, GuideGPT was compared with a generic version of the same model (PureGPT) across 30 questions about medication for osteonecrosis of the jaw. Ten international experts evaluated responses for content, language, scientific explanation, and agreement using 5-point Likert scales, with statistical testing. GuideGPT significantly outperformed the generic model in content, scientific explanation, and agreement, confirming the benefit of RAG in enhancing accuracy and transparency. The authors noted limitations including occasional inconsistency in language quality, dependence on available literature quality, and potential bias when guideline sources contained conflicting information.

Together, these studies demonstrate the growing use of LLM- and RAG-based systems to enhance medication communication, factual grounding, and patient comprehension across languages and clinical domains. However, most employed earlier-generation models such as GPT-3.5 or Mistral-7B, with limited evaluation against newer models such as GPT 5.1. Several integrated external medical knowledge, but many lacked standardized evaluation procedures and transparent reporting. Importantly, the literature has largely overlooked CBD-focused education and the specific needs of older adults, including those with mild cognitive impairment. Moreover, there is currently no publicly available benchmark dataset for CBD-related education [20], [21], [22], [23], nor any standardized metric for evaluating LLM and RAG systems specifically on medication education problems. Existing evaluations are often inconsistent and rely heavily on subjective human scoring rather than scalable, reproducible methods. The present study addresses these gaps by systematically evaluating advanced LLM and RAG architectures for safe, context-aware CBD communication tailored to vulnerable aging populations and individuals with cognitive impairment, while introducing annotation-free evaluation methods to enable scalable, objective assessment of model performance.

**Methods**
This section describes the methodological framework developed for evaluating the ability of LLMs and RAG systems to provide evidence-grounded educational information on CBD use for older adults. The framework includes the generation of structured scenarios that serve as standardized inputs, the design of prompt engineering techniques for both standalone models and retrieval-based systems, the configuration of the models and external evidence used in the retrieval pipeline, and the automated approaches employed to evaluate model outputs. Together, these components establish a controlled and reproducible environment for assessing the reliability of model-generated educational content.

*Parametric Scenario Generation and Input Construction*
In alignment with previous work on automated scenario synthesis and parametric prompt generation for language model evaluation [37], [38], [39], [40], this study employed a structured and reproducible framework to design representative cases for systematic model assessment. The objective was to create diverse and realistic depictions of older adults seeking educational information about CBD use for symptom management while maintaining control over the variation of clinical and contextual parameters.

A total of sixty-four distinct scenarios were developed following a balanced factorial design 4 × 4 × 4, covering combinations of three conceptual axes: symptom goal, preferred form of administration, and cognitive status. The symptom goals included pain, sleep disturbance, anxiety, and other general well-being concerns [41]. The forms of administration comprised oral oil, capsule, topical application, and no specific preference [7], [41]. Cognitive status was varied across four categories: none, subjective cognitive impairment, mild cognitive impairment, and cognitive impairment [42], [43].

Each scenario integrated variations in demographic and health attributes, including sex, age (65–85 years), height (155–185 cm), body weight (50–90 kg) [44], [45], and hepatic and renal function (normal or mild impairment) [46]. To increase ecological validity, additional diversity was introduced through comorbidities (such as mild hypertension, type 2 diabetes, and osteoarthritis) and common medication regimens (for example, metformin, acetaminophen, or vitamin D supplementation) [1], [47], [48]. Each case also specified the availability of caregiver support, alternating between older adults living independently and those receiving informal assistance [49]. The cannabis history attribute identified whether the individual was CBD-naive or THC-sensitive [12].

This scenario generation approach ensured comprehensive coverage of user characteristics while preserving reproducibility and interpretability. It provided a consistent set of semantically rich, clinically relevant inputs for evaluating both LLMs and RAG systems under controlled, comparable conditions.

*Prompt Engineering and Adaptation*
Prompt engineering refers to the careful design of instructions, contextual details, and user inputs given to an LLM to shape how it interprets a query and generates a response. It allows developers and users to guide an LLM's reasoning process by specifying what information it should consider, how it should structure its output, and which safety constraints it must follow [50]. In this study, prompt engineering followed established frameworks for structured and safety-aware language model design in medical and educational applications [28], [50]. The process emphasized clarity, contextual grounding, and transparent reasoning to support factual accuracy and ethical alignment. Two complementary prompt configurations were developed, one for standalone LLMs operating independently (Figure 1 (a)), and another for RAG systems incorporating external resources (Figures 1 (b) and (c)). Each configuration applied multiple prompt engineering techniques to promote interpretability and reproducibility.

The baseline LLM prompt applied a combination of role prompting [51], schema-guided output prompting [52], and chain-of-thought prompting [53], [54]. Role prompting instructed the model to act as an educational assistant that organizes user-provided information about CBD and produces factual, neutral, and non-prescriptive responses. Schema-guided output prompting defined a fixed response structure comprising fields for reasoning summary, dosage education, titration steps, monitoring guidance, and a disclaimer. Chain-of-thought prompting operationalized the model's reasoning process through four defined pathways, decline, inform, educate, and uncertain, ensuring consistent behavior based on the completeness and safety of each scenario. The prompt was refined iteratively until it achieved stable, reproducible, and semantically coherent outputs across model families. The above described system prompt used to standardize standalone LLM behavior is provided in Multimedia Appendix 1, and the representative scenario, described in the previous subsection, that serves as the human prompt is provided in Multimedia Appendix 2.

For the RAG configuration, the LLM prompt was adapted to incorporate retrieval-grounded prompting along with evidence-constrained chain-of-thought reasoning [28], [55]. The model was instructed to base its reasoning primarily on retrieved document content. This adaptation positioned the model as an evidence-informed reasoning system, improving factual transparency and reducing unsupported generation. An additional output field captured document identifiers and excerpts corresponding to each cited source to support verification and traceability. The adapted system prompt for the RAG configuration is provided in Multimedia Appendix 3, and the human prompt followed the same scenario structure shown in Multimedia Appendix 2.

This combined prompt framework, described in Multimedia Appendices 1 through 3, established a controlled environment that supported explainability, consistency, and adherence to safety and educational standards. Restricting the input to a predefined schema functioned as a safety safeguard by preventing the inclusion of unnecessary or identifiable personal information, and schema-guided output prompting ensured a standardized response format that reduced harmful content generation, minimized hallucinations, and enabled fair comparison across LLM and RAG configurations.

*Model Configuration and Retrieval-Augmented Pipeline*
This study evaluated multiple LLMs and their RAG counterparts using identical inputs and prompt structures to ensure reproducibility and comparability. The models tested included OpenAI GPT 5.1 [25], Google Gemini 2.5 Pro [26], Mistral AI Medium 3 [56], Anthropic Claude Sonnet 4.5 [27], xAI Grok 4 [57], and DeepSeek V3.2-Exp [58]. These models were selected as the most advanced publicly available systems as of December 2025 and accessed through official application programming interfaces (APIs). All models were configured with consistent inference parameters, low temperature values to reduce stochasticity [59] and fixed output token limits [60] to prevent truncation or excessive length.

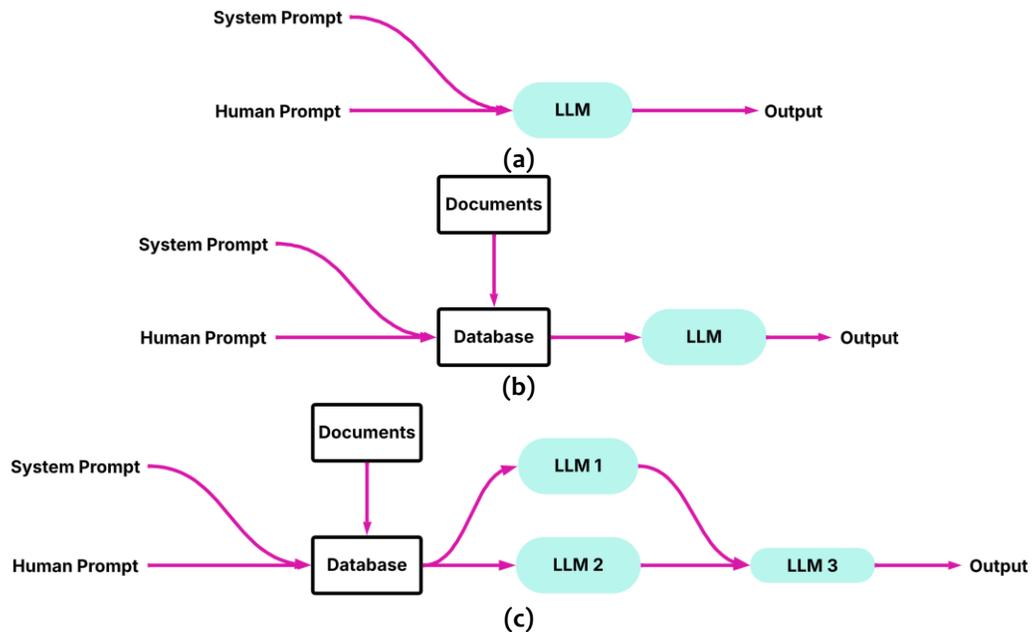

**Figure 1.** The block diagram of (a) a standalone large language model (LLM) receiving a human prompt together with a system prompt that instructs how the LLM should interpret the human input and generate its response, (b) the standard retrieval-augmented generation configuration in which an LLM is equipped with retrieved documents as external resources, and (c) the advanced configuration where two distinct retrieval-augmented LLMs (LLM 1 and LLM 2) generate responses that are evaluated by a third LLM acting as the tiebreaker that generates the final output.

For the RAG configuration, Figure 1 (b), each base model was integrated into a retrieval pipeline implemented with the LangChain framework [61]. A total of thirty-two curated educational and clinical resources were used as the external evidence base, including Canadian and international guidelines on CBD use among older adults and individuals with cognitive impairment, as well as brochures and educational webpages from reputable health organizations. The list of documents included in the retrieval corpus is provided in Multimedia Appendix 4.

All documents were preprocessed and converted into vector embeddings to enable efficient similarity search and contextual retrieval. The system used the Facebook AI Similarity Search (FAISS) [62] for indexing and semantic retrieval of relevant educational and pharmacological materials related to CBD and cannabis use in older adults. Source documents were divided into overlapping text chunks of fixed length to preserve semantic continuity across boundaries [61]. A predefined number of top-ranked chunks, selected by cosine similarity to the query, were appended to the model prompt as contextual evidence.

In addition to the standard RAG configuration, referred to as the naive configuration and shown in Figure 1 (b), an extended configuration was also implemented as shown in Figure 1 (c). In the advanced setup, two distinct LLMs (LLM

1 and LLM 2) each receive the user input together with the retrieved documents and generate individual responses. A third tiebreaker LLM then reviews both outputs along with the original user input and the retrieved evidence and selects or generates the final answer according to the prompt provided in Multimedia Appendix 5.

*Annotator-Free Evaluation of LLM and RAG System Performance*
Publicly available or validated ground truth datasets for educational CBD communication are currently lacking [63], [64], [65]. In addition, human and expert annotation introduce several limitations, including subjectivity, variability across evaluators with differing clinical or psychiatric perspectives [66], [67], and the substantial time and financial cost required to generate high-quality labeled data. To address these constraints, this study employed an automated, expert annotation-free evaluation framework. Grounded in the existing literature on automated evaluation [68], [69], [70], three quantitative, human-free evaluation methods were designed and implemented to assess model performance across complementary dimensions: statistical consensus evaluation, feature-aligned directional evaluation, and LLM-as-a-judge rubric evaluation. Together, these approaches enabled a comprehensive and model-unbiased assessment of logical coherence, safety compliance, and dosage characteristics without relying on human annotation.

Statistical Consensus Evaluation
This method evaluated how closely each model's dosage and titration outputs aligned with the collective pattern produced across all models. For every input scenario, the mean and variability of dosage and titration parameters were computed across the full model set, and each model's deviation from this distribution was quantified using standardized z-scores. Positive deviations reflected relatively aggressive dosing tendencies, while negative deviations reflected more cautious dosing tendencies.

By anchoring evaluation to the distribution formed by all participating models, this approach enabled comparison even in the absence of a gold standard [71], [72], [73]. The aggregate behavior of diverse models developed by different companies and research groups functioned as an implicit benchmark. Similar consensus-based evaluation strategies have been used in prior work by treating multiple heterogeneous models as reference distributions for assessing reliability and robustness [71], [72], [73].

Feature-Aligned Directional Evaluation
This method evaluated whether a model's dosage outputs changed in clinically expected directions based on scenario characteristics. Older age, CBD-naive status, renal or hepatic impairment, limited caregiver support, THC sensitivity, and cognitive impairment are factors that generally call for more cautious educational dosing [12], [74], [75]. Each scenario was assigned an expected directional score based on these attributes, and a model's output was classified as aligned, misaligned, or neutral depending on whether its dosage shifted in the anticipated direction. The resulting alignment rates offered insight into each model's safety awareness and its sensitivity to clinically vulnerable profiles.

LLM-as-a-Judge Rubric Evaluation

In this method, an advanced language model independently evaluated the quality of each model's output using a structured scoring rubric [76]. The rubric included five quality dimensions: relevance to the user's stated goal, factual grounding, safety of educational communication, adherence to the required response structure, and clarity for lay readers. For each dimension, the evaluating model assigned a sub-score ranging from 0 (low) to 5 (high) and also generated a weighted total score on the same 0 to 5 scale using a proportional weighting scheme applied to the five sub-scores. This enabled consistent and comparable quality assessment across systems. The approach drew on growing evidence that language models can serve as reliable evaluators, approximating expert review while avoiding the variability, cost, and limited scalability associated with human annotation [72], [73], [76], [77].

**Results**

This section summarizes the performance of the proposed methodology across all generated scenarios and reports the results obtained from the automated evaluation methods.

Among the models tested, Anthropic Claude Sonnet 4.5 [27] and related variants such as Claude Haiku 4.5 did not generate educational CBD content. Instead, they returned messages such as "Even with excellent safeguards in the prompt, I am not comfortable acting as a systematic intake processing tool for cannabis or CBD guidance because the framing creates clinical-looking output, I cannot verify safety in practice, and systematic processing bypasses appropriate caution…" Such responses reflect a conservative, safety-oriented design and may still be informative in real-world settings. However, because they did not produce the structured educational outputs required by the evaluation framework, Claude Sonnet 4.5 [27] was excluded from quantitative analysis.

Due to their superior performance on major LLM benchmarks [78] as of December 2025, GPT 5.1 [25] and Gemini 2.5 Pro [26] were selected as the base LLMs for the RAG configurations. Two different standard RAG configurations, shown in Figure 1 (b), were implemented and evaluated separately using GPT 5.1 and Gemini 2.5 Pro. For the ensemble RAG configuration, shown in Figure 1 (c), GPT 5.1 and Gemini 2.5 Pro served as LLM 1 and LLM 2, and an additional separate instance of GPT 5.1 was used as the third tiebreaker LLM. The temperature for both models was set to 0 to produce more deterministic outputs [59]. The optimal chunk size and chunk overlap were 1200 and 150 tokens, respectively, and for each query, the top $k = 6$ retrieved document chunks were supplied as evidence during RAG inference [24].

Figures 2 (a) through 2 (e) show boxplots for five CBD-related parameters, including dosage (mg), dosing frequency per day, titration amount (mg), titration interval (days), and maximum daily dose (mg), generated by the evaluated models. DeepSeek V3.2-Exp, Grok 4, and Medium 3 produced the highest CBD dosage, titration amounts, and maximum daily dose recommendations, reflecting more permissive and variable dosing behavior. In contrast, GPT 5.1, Gemini 2.5 Pro, and the RAG-based models generated much lower and more narrowly distributed dosage and titration values, typically in the range of 2 to 5 mg, indicating more cautious and consistent recommendations. A similar pattern was observed for dosing frequency. Although the RAG-based systems recommended shorter titration intervals, their maximum daily doses remained substantially lower than those produced by the standalone LLMs.

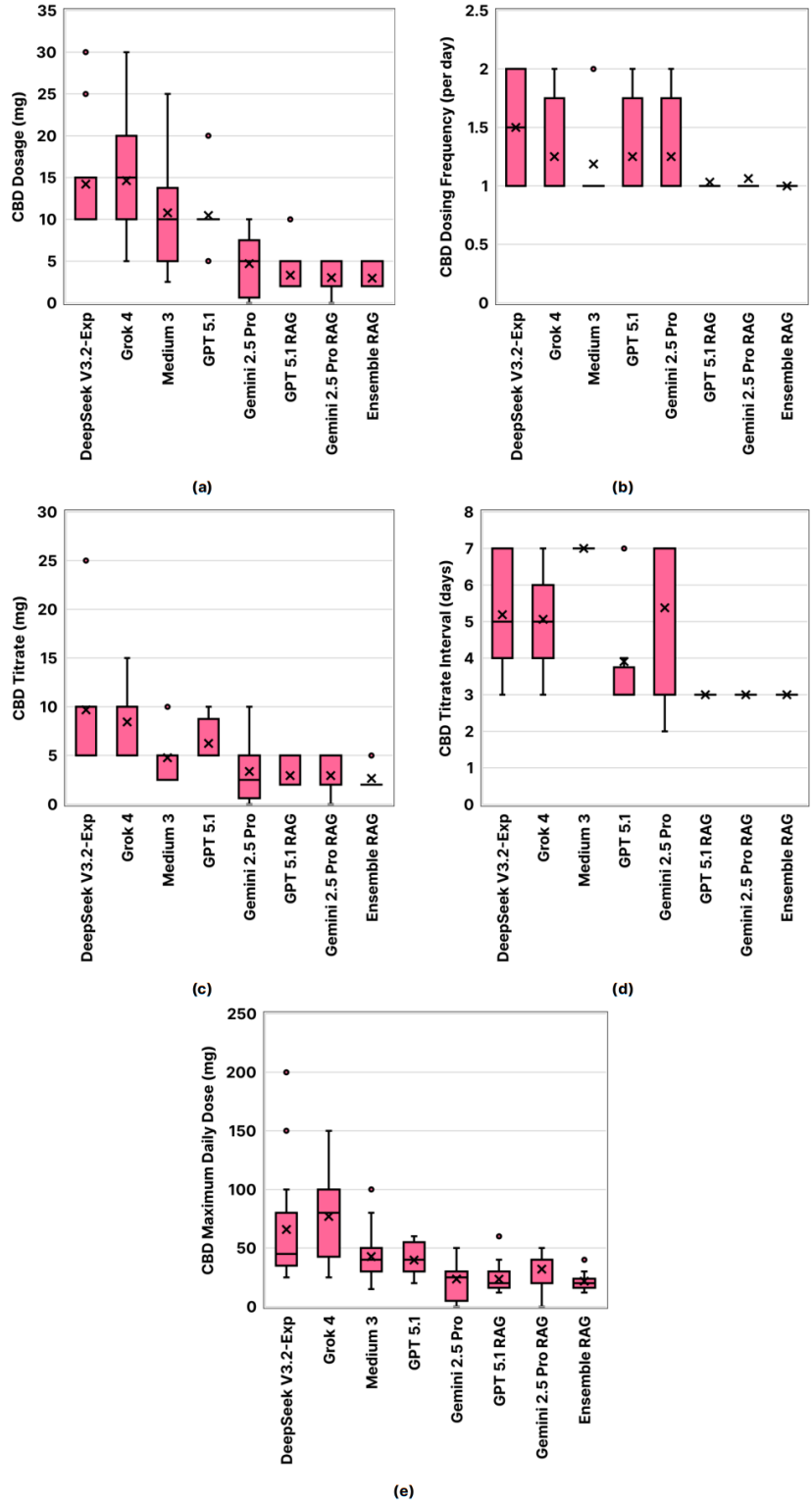

**Figure 2.** Boxplots of the generated educational content on CBD (a) dosage in milligrams, (b) dosing frequency per day, (c) titration amount in milligrams, (d) titration interval in days, and (e) maximum daily dose in milligrams across the evaluated LLM and RAG systems.

As described in the prompt engineering subsection, the generated scenarios serving as the human prompt, an example of which is illustrated in Multimedia Appendix 2, were combined with the system prompt for the retrieval-augmented models provided in Multimedia Appendix 3 to form the full input to the ensemble RAG system. For illustration, the structured output produced by the ensemble RAG configuration in response to this input can be viewed in Multimedia Appendix 6.

*Statistical Consensus Evaluation*
Figures 3 (a) through 3 (e) present the mean and standard deviation of the generated values, along with the corresponding standardized z-scores, for five CBD-related parameters. The z-scores were derived from the statistical consensus evaluation method.

Examining the standardized z-scores in the last two columns of Figures 3(a)-(e), the RAG-based models exhibited a distinct pattern compared with the standalone LLMs. They showed negative mean z-scores and relatively low standard deviations, indicating that their outputs were consistently more cautious than the overall model average and varied little across scenarios. This pattern suggests that the RAG-based models adjusted their recommendations more conservatively in response to scenario-specific factors, producing lower values that reflect the influence of external clinical CBD usage evidence.

In comparison, the standalone GPT 5.1 and Gemini 2.5 Pro models showed more moderate behavior, with mean z-scores close to zero and higher standard deviations. This pattern indicates strong alignment with the central tendency of all model outputs while still allowing for variability across scenarios, suggesting that GPT 5.1 and Gemini 2.5 Pro generated balanced, predictable values that remained responsive to scenario-specific differences. DeepSeek V3.2-Exp and Grok 4 showed positive mean z-scores, indicating consistently higher outputs than the consensus. DeepSeek exhibited the largest overall deviations, whereas Grok demonstrated more moderate divergence but maintained an upward bias, particularly in titration dose. Medium 3 was comparatively more conservative, with lower mean z-scores and smaller deviations across parameters.

Together, these results indicate that the RAG-based models offered the most conservative behavior while remaining closely aligned with the consensus, reflecting their grounding in external clinical guidelines on CBD use. GPT 5.1 and Gemini 2.5 Pro represented the most stable and balanced models overall, whereas DeepSeek V3.2-Exp and Grok 4 produced higher outputs and showed greater divergence from the consensus.

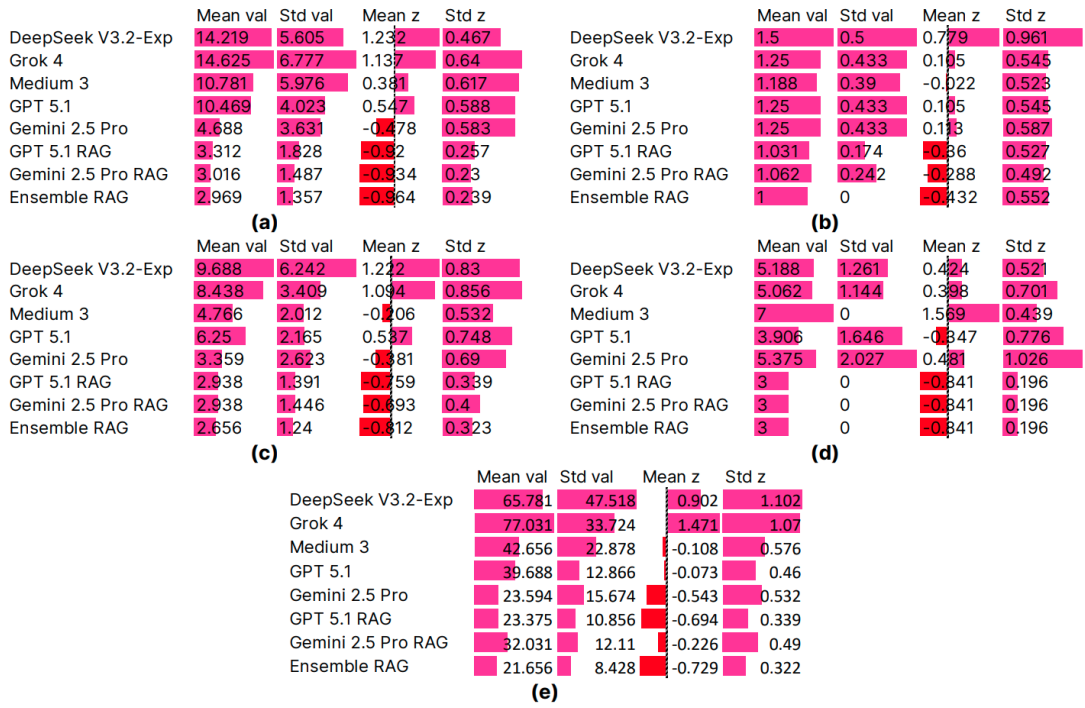

**Figure 3.** The mean and standard deviation of the generated values (Mean val and Std val) as well as statistical consensus evaluation showing standardized z-scores (Mean z and Std z) of CBD (a) dosage in milligrams, (b) dosing frequency per day, (c) titration amount in milligrams, (d) titration interval in days, and (e) maximum daily dose in milligrams across the evaluated LLM and RAG systems.

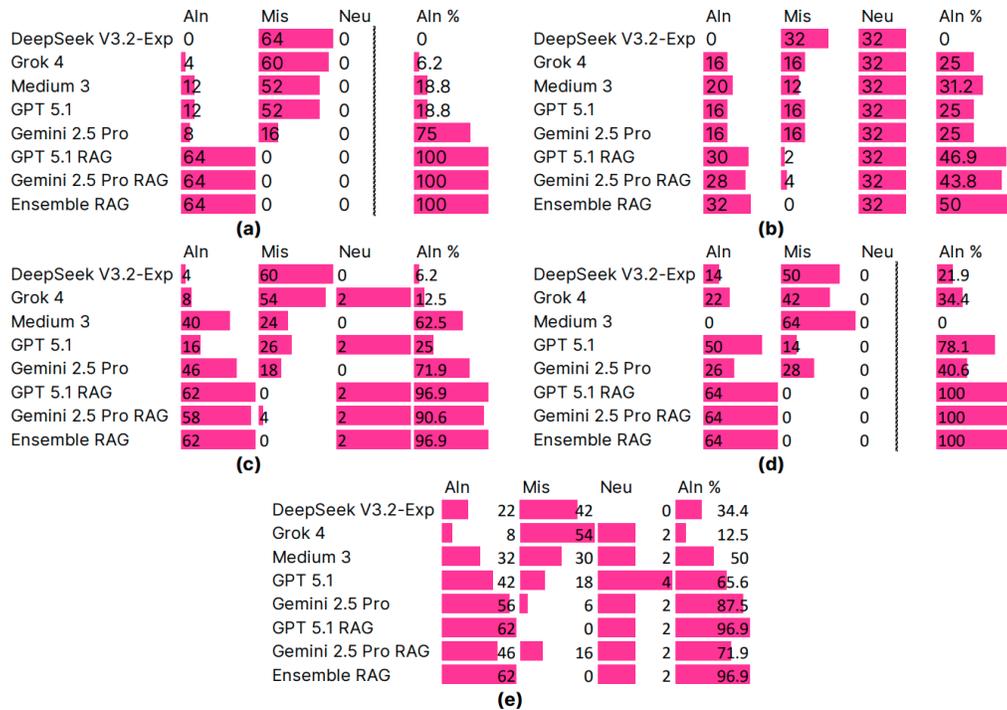

**Figure 4.** Feature-aligned directional evaluation showing the number of aligned (Aln), misaligned (Mis), and neutral (Neu) outputs, along with alignment rates (Aln%) for the generated CBD (a) dosage in milligrams, (b) dosing frequency per day, (c) titration amount in milligrams, (d) titration interval in days, and (e) maximum daily dose in milligrams across the evaluated LLM and RAG systems.

*Feature-Aligned Directional Evaluation*

Figures 4 (a) through 4 (e) present the number of aligned, misaligned, and neutral outputs, along with alignment rates for the generated CBD dosage and titration parameters. The feature-aligned directional evaluation revealed clear differences in how the models adjusted their outputs in response to demographic and clinical risk factors. The RAG-based models, with the ensemble RAG performing the best, achieved the highest alignment rates, indicating that they consistently produced more cautious educational values when intake features such as older age, CBD-naive status, THC sensitivity, or renal and hepatic impairment were present. Nevertheless, alignment rates were lowest for CBD dosing frequency, likely because this parameter has a limited set of possible values, typically once or twice per day. The standalone GPT 5.1 and Gemini 2.5 Pro models showed lower alignment than their retrieval-augmented counterparts, underscoring the benefit of incorporating external knowledge into LLMs. In contrast, DeepSeek V3.2-Exp and Grok 4 showed limited alignment with expected directional patterns, often shifting outputs in ways that did not reflect cautious guidance for vulnerable older adults. Medium 3 performed considerably better, demonstrating moderate sensitivity to risk-related features such as age, cognitive status, comorbidity burden, polypharmacy, and hepatic or renal impairment.

*LLM-as-a-Judge Rubric Evaluation*

Figures 5 (a) and 5 (b) present the LLM-as-a-Judge rubric evaluation scores, with GPT 5.1 and Gemini 2.5 Pro separately serving as the evaluators. The rubric-based evaluation revealed a consistent pattern across both judging models. When GPT 5.1 served as the evaluator, the highest-scoring systems were the RAG-based configurations, with GPT 5.1 RAG and the Ensemble RAG achieving total scores above 4.36, followed closely by Gemini 2.5 Pro RAG. The standalone frontier models, GPT 5.1 and Gemini 2.5 Pro, formed a clear middle tier with total scores around 4.0 to 4.1, reflecting strong performance but not matching the added structure and evidence grounding provided by retrieval augmentation. DeepSeek V3.2-Exp, Grok 4, and Medium 3 received notably lower safety and clarity scores under this judge, resulting in total scores between 3.5 and 4.0, with DeepSeek showing the strongest downward pull due to lower safety evaluations.

When Gemini 2.5 Pro acted as the judge, the relative ranking of models remained stable, indicating high agreement between evaluators. Gemini 2.5 Pro and its RAG variant achieved the highest overall scores, reaching totals near 4.9 and 4.8, respectively, while the Ensemble RAG closely followed at 4.76. GPT 5.1 and GPT 5.1 RAG maintained strong scores above 4.6, again forming a consistent middle tier. DeepSeek V3.2-Exp and Grok 4 remained the lowest-scoring models, reflecting persistent weaknesses in safety and structure dimensions across both judges. Medium 3 demonstrated moderate performance similar to its placement under the GPT 5.1 judge. Together, these patterns indicate robust agreement across evaluation models, with retrieval-augmented systems consistently producing the highest-quality educational outputs, while some standalone models exhibited greater challenges, particularly in safety-oriented aspects of the rubric.

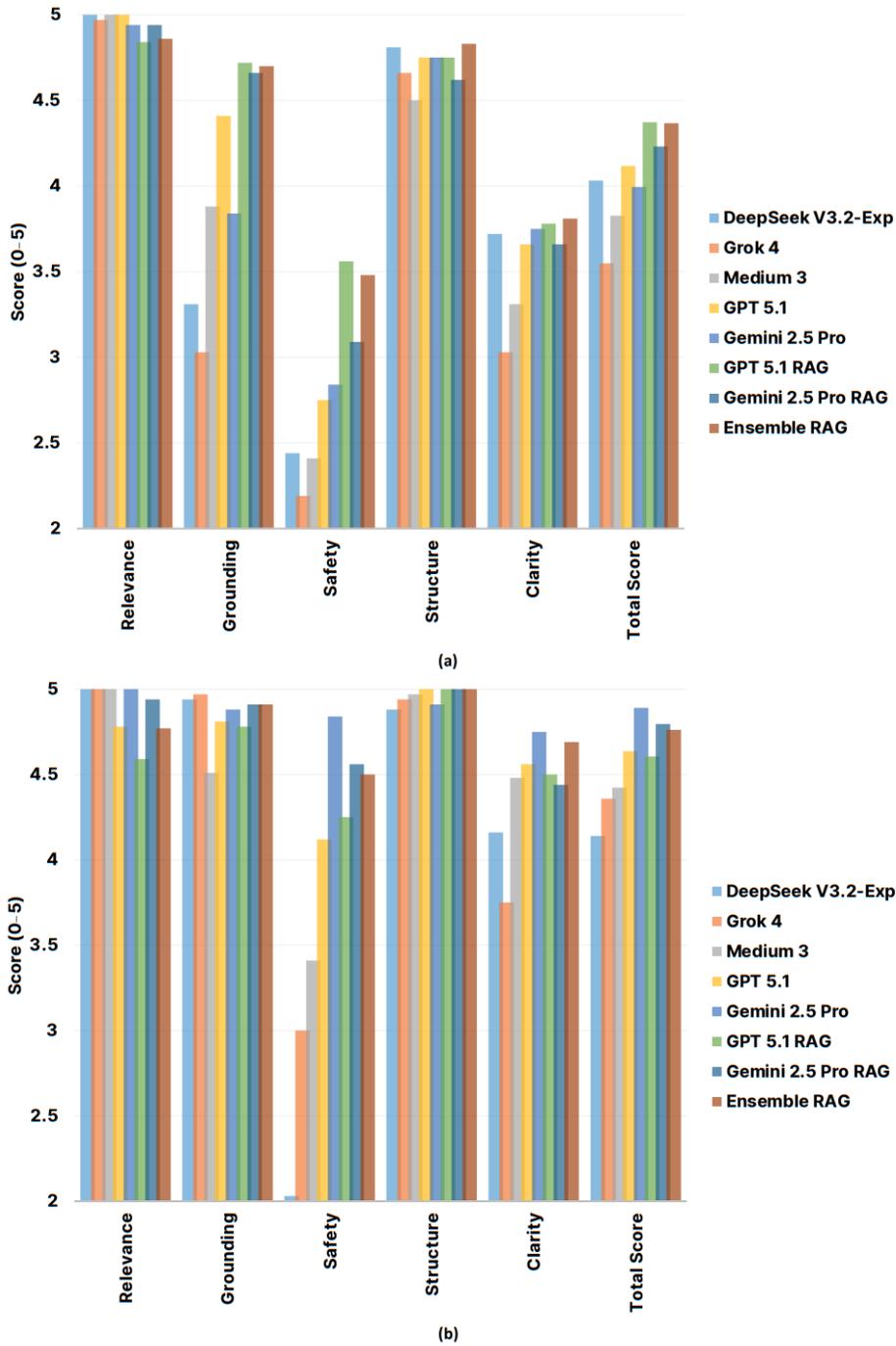

**Figure 5.** LLM-as-a-judge rubric-based evaluation of model outputs across five quality dimensions and the total score, using (a) GPT 5.1 and (b) Gemini 2.5 Pro as the evaluating models.

## Discussion

This study developed a prompt-engineered, retrieval-augmented framework for CBD education tailored to older adults, including those with cognitive impairment, and introduced an automated, annotator-free evaluation pipeline for assessing the performance of LLM and RAG systems in this context. Across diverse intake profiles, retrieval-augmented systems demonstrated consistently strong performance, producing more coherent and more contextually responsive outputs than standalone models. Their use of retrieved evidence enabled adaptive guidance aligned with age-related and comorbidity-related factors. Standalone Gemini 2.5 Pro and GPT 5.1 performed well with strong grounding, consistency, and clarity, but were less consistently cautious without retrieval support. DeepSeek V3.2-Exp and Grok 4 generated higher and more variable dosage values and showed limited sensitivity to risk-related features such as age, cognitive status, comorbidity burden, polypharmacy, and hepatic or renal impairment. Overall, the automated evaluation clearly differentiated model behavior and confirmed the benefit of retrieval grounding for safety and alignment.

The findings align with the broader literature on LLM safety and RAG grounding, which consistently shows that access to relevant external evidence improves factual reliability and reduces unsafe extrapolation [24]. Existing evaluation frameworks, however, often rely on human annotation or gold standard labels that are expensive, subjective, and impractical for emerging domains such as CBD education [66]. This study addresses this gap by demonstrating that the proposed statistical consensus metrics, directional feature alignment, and LLM-based rubric scoring can serve as scalable alternatives that preserve consistency while reducing reliance on resource-intensive human review [76].

This work lays the foundation for more cautious and context-aware educational agents that help older adults and caregivers make informed decisions about CBD use. The evaluation framework offers a scalable alternative to expert annotation, enabling rapid assessment of new LLM and RAG systems in sensitive health domains. Integrating structured retrieval with feature-aware reasoning can enhance transparency, strengthen safety, and support trust in AI-supported education across clinical and community settings.

The study design combines controlled scenario generation, standardized prompting, and complementary automated evaluation to enable fair, reproducible comparison across models. A 4 × 4 × 4 factorial grid yields 64 intake cases that vary in symptom goal, administration form, and cognitive status while incorporating key clinical attributes. Parallel prompts for standalone and retrieval-augmented models use fixed schemas and low temperature for consistency. A curated corpus of 32 CBD guidelines and education resources grounds retrieval. An ensemble configuration with adjudication improves robustness and caution. Three automated evaluations, consensus statistics, feature-aligned directional checks, and rubric-based LLM judging, provide convergent evidence with transparent, source-traceable outputs.

This study has several limitations. The retrieval-augmented approach was intentionally simple, relying on fixed prompt engineering and basic evidence injection rather than adaptive retrieval, hierarchical selection, or domain-specific retrievers [79],

[80]. The evaluation pipeline did not include targeted human expert review to validate or calibrate automated scores [66], [76]. In addition, portions of the evaluation relied on LLM-based judges, which may introduce model-specific biases, as models appeared to assign relatively higher scores to outputs generated by closely related architectures. The scenario space, while structured, did not fully capture multimorbidity patterns, polypharmacy complexity, or varied caregiver roles that occur in practice. The analysis focused on single-turn responses and did not assess multi-turn dialogue, error recovery, or sustained safety. The study relied on a limited set of curated CBD-related educational resources, which may constrain the breadth of evidence available for reasoning. Finally, the work operated without gold standard datasets for CBD education and did not fine-tune models on domain-specific corpora, which may limit alignment with clinical expectations. Additionally, some models, such as Claude Sonnet 4.5, produced conservative refusal responses that may be informative in real-world contexts but were excluded from quantitative analysis due to the predefined evaluation criteria, highlighting ambiguity between usefulness and evaluability within the current framework.

To address the aforementioned limitations, future work should adopt advanced retrieval-augmented designs that use adaptive retrieval, hierarchical evidence selection, and domain-specific retrievers to improve grounding in complex cases. Targeted human expert review should be integrated to validate and calibrate automated scores and to provide clinical oversight. The scenario space should be expanded to reflect richer multimorbidity patterns, polypharmacy, and caregiver dynamics for greater ecological validity. Importantly, older adults should be meaningfully engaged in future studies to contribute their perspectives, information needs, and lived experiences to system design and evaluation. Multi-turn dialogue should be evaluated to examine robustness, error recovery, and sustained safety over longer interactions. When reliable gold standard datasets become available, models should be fine-tuned on domain-specific CBD education corpora and benchmarked against external datasets to assess generalization across settings.

Educational guidance on CBD for older adults must prioritize non-prescriptive language, clear disclaimers, and encouragement to consult clinicians for individualized care. Retrieval grounding should include transparent citation of sources and traceable excerpts to support verification. Privacy safeguards are essential when handling intake attributes such as age, comorbidities, and medication lists, with strict exclusion of identifiable data from prompts and outputs. Accessibility should be ensured through plain language, readable formatting, and caregiver-aware messaging for users with cognitive impairment. Bias audits are needed to check for systematic differences across sex, age bands, organ function, and multimorbidity patterns. Human oversight remains necessary for edge cases, conflicts between guidelines, and potential adverse event signals. Ongoing maintenance is also required, including routine updating of evidence sources and monitoring for knowledge drift to preserve clinical relevance over time.

This work delivers a retrieval-augmented framework and an annotator-free evaluation approach that together advance safe, transparent CBD education for older adults. The results show consistent benefits from structured retrieval and feature-aware reasoning, with ensemble adjudication yielding the most reliable outputs. Future extensions to adaptive retrieval, richer scenarios, and selective expert review can strengthen clinical relevance and support responsible deployment across sensitive health domains.

**Acknowledgements**
This research was supported by the Alzheimer's Association, whose contribution is gratefully acknowledged.

**Funding**
This research was funded by the Alzheimer's Association through a grant awarded to Charlene H. Chu.

**Conflicts of Interest**
The authors report no conflicts of interest.

**Data Availability**
All data and code are publicly available on GitHub.

**Authors' Contributions**
Ali Abedi, Charlene H. Chu, and Shehroz S. Khan conceived the study. Ali Abedi designed the study setting and methodology, generated the data, designed and implemented the algorithms, developed analysis code, produced the results, and drafted all sections of the manuscript. All authors reviewed and approved the final manuscript.

**Abbreviations**
AI: artificial intelligence
CBD: cannabidiol
LLM: large language models
RAG: retrieval-augmented generation
THC: tetrahydrocannabinol